\documentclass[12pt]{article}
\usepackage{amsmath,amsfonts,amssymb}

\makeatletter
\def\numberbysection{\@addtoreset{equation}{section}
    \def\theequation{\thesection.\arabic{equation}}}
\makeatother

\numberbysection

\newcommand{\be}{\begin{eqnarray}}
\newcommand{\ee}{\end{eqnarray}}
\newcommand{\non}{\nonumber}
\newcommand{\id}{\mathbb{I}}
\newcommand{\tr}{\mathop{\rm tr}\nolimits}
\newcommand{\str}{\mathop{\rm str}\nolimits}

\begin{document}

\begin{titlepage}
\strut\hfill UMTG--258
\vspace{.5in}
\begin{center}

\LARGE ${\cal N}=6$ super Chern-Simons theory $S$-matrix\\
and all-loop Bethe ansatz equations\\
\vspace{1in}
\large Changrim Ahn \footnote{
       Department of Physics, Ewha Womans University,
       Seoul 120-750, South Korea} and
       Rafael I. Nepomechie \footnote{
       Physics Department, P.O. Box 248046, University of Miami,
       Coral Gables, FL 33124 USA}\\

\end{center}

\vspace{.5in}

\begin{abstract}
We propose the exact $S$-matrix for the planar limit of the ${\cal
N}=6$ super Chern-Simons theory recently proposed by Aharony, Bergman,
Jafferis, and Maldacena for the ${\rm AdS}_4/{\rm CFT}_3$
correspondence.  Assuming $SU(2|2)$ symmetry, factorizability and
certain crossing-unitarity relations, we find the $S$-matrix including
the dressing phase.  We use this $S$-matrix to formulate the
asymptotic Bethe ansatz.  Our result for the Bethe-Yang equations and
corresponding Bethe ansatz equations confirms the all-loop Bethe
ansatz equations recently conjectured by Gromov and Vieira.
\end{abstract}

\end{titlepage}

\setcounter{footnote}{0}

\section{Introduction}\label{sec:intro}

The AdS/CFT correspondence \cite{Maldacena,GKP,Witten}, which has led
to many exciting developments in the duality between type IIB string
theory on $AdS_5\times S^5$ and ${\cal N}=4$ super Yang-Mills (YM) theory,
is now being extended into $AdS_4/CFT_3$ \cite{CFT3}.  A most
promising candidate is ${\cal N}=6$ super Chern-Simons (CS) theory
with $SU(N)\times SU(N)$ gauge symmetry and level $k$.  This model,
which was first proposed by Aharony, Bergman, Jafferis, and Maldacena
\cite{ABJM}, is believed to be dual to M-theory on $AdS_4\times
S^7/Z_k$.  Furthermore, in the planar limit of $N, k\to\infty$ with a
fixed value of 't Hooft coupling $\lambda=N/k$, the ${\cal N}=6$ CS is
believed to be dual to type IIA superstring theory on $AdS_4\times
CP^3$.  This model contains two sets of scalar fields transforming in
bifundamental representations of $SU(N)\times SU(N)$ along with
respective superpartner fermions and non-dynamic CS gauge
fields.  (For some subsequent developments, see \cite{CSref, NT}.)

The integrability of the planar ${\cal N}=6$ CS was first discovered by
Minahan and Zarembo \cite{MZ} in the leading two-loop-order
perturbative computation of the anomalous dimensions of
gauge-invariant composite operators.  They found that the
dilatation operator for the scalar operators is an integrable
Hamiltonian of an $SU(4)$ spin chain with sites alternating between
the fundamental and anti-fundamental representations.  They obtained
corresponding two-loop Bethe ansatz equations (BAEs) using algebraic
Bethe ansatz results for an inhomogeneous spin chain with different
representations developed in \cite{KulRes}.  They then conjectured
two-loop BAEs for all operators (including the fermions) corresponding
to the full $OSp(2,2|6)$ superconformal group .

More recently, all-loop BAEs for the ${\cal N}=6$ CS were conjectured
by Gromov and Vieira \cite{GV} based on the perturbative result \cite{MZ}
and the classical integrability in the large-coupling limit
discovered in \cite{AruFro,Stef,GroVieii}.

The purpose of this note is to propose an exact $S$-matrix for the
${\cal N}=6$ CS, and to derive the all-loop BAEs \cite{GV} from this
$S$-matrix.  A factorizable $S$-matrix has played an important role in
the developments of $AdS_5/CFT_4$.  Indeed, an $S$-matrix describing
the scattering of excitations of the dynamic spin chain corresponding
to planar ${\cal N}=4$ YM has been proposed by Beisert
\cite{Be}, and a related $S$-matrix describing the scattering of
world-sheet excitations of the $AdS_{5}\times S^{5}$ superstring sigma
model has been proposed by Arutyunov, Frolov and Zamaklar (AFZ)
\cite{AFZ}.  These $S$-matrices have been derived from the assumption
that the excitations are described by a Zamolodchikov-Faddeev (ZF)
algebra \cite{ZZ, Fa}, and that they have a centrally extended
$su(2|2) \oplus su(2|2)$ symmetry \cite{Be}.  The AFZ ``string''
$S$-matrix obeys the standard Yang-Baxter equation, while Beisert's
$S$-matrix obeys a twisted (dynamical) Yang-Baxter equation.  The
$S$-matrices are a very useful tool to overcome shortcomings from the
Bethe ansatz approach such as the wrapping problem \cite{BajJan} and
for full quantization of the string theory.

As discussed in more detail below, ${\cal N}=6$ CS has two sets of
excitations, namely $A$-particles and $B$-particles, each of which
form a four-dimensional representation of $SU(2|2)$.
We propose an $S$-matrix with the following structure: \footnote{This
structure is similar to that of the $S$-matrix proposed in \cite{ZZ2}
for the $O(3)$ sigma model with $\theta=\pi$, where the two types
of particles are the left-movers and right-movers.}
\be
S^{AA}(p_1,p_2)&=&S^{BB}(p_1,p_2)=S_0(p_1,p_2){\widehat S}(p_1,p_2)
\,, \non \\
S^{AB}(p_1,p_2)&=&S^{BA}(p_1,p_2)={\tilde S}_0(p_1,p_2){\widehat
S}(p_1,p_2) \,, \non
\ee
where $\widehat S$ is the matrix part determined by the $SU(2|2)$
symmetry, and is essentially the same as that found for ${\cal N}=4$ YM in \cite{AFZ}.
An important difference arises in the dressing phases $S_0, {\tilde
S}_0$ due to the fact that the
$A$- and $B$-particles are related by complex conjugation.
Crossing symmetry \cite{Ja} relates the two phases and gives different
dressing phases for the $S$-matrices.
These dressing phases play a crucial role in the process of deriving the Bethe-Yang equations.

The outline of this paper is as follows.  In Section 2, we
use the bulk ZF algebra and crossing relations to construct the $S$-matrix.
In Section 3, all-loop asymptotic BAEs are derived
from diagonalizing the Bethe-Yang matrix.
We conclude in Section 4 with a brief discussion of our results.

\section{Excitations and $S$-matrix}\label{sec:Smatrix}

We recall \cite{ABJM} that the ${\cal N}=6$
CS theory has a pair of
scalar fields $A_{i}$ ($i = 1, 2$) in the bifundamental representation
$({\bf N}, {\bf \bar N})$ of the $SU(N) \times SU(N)$ gauge group, and another pair of scalar fields $B_{i}$
($i = 1, 2$) in the conjugate representation $({\bf \bar N}, {\bf
N})$. The vacuum is given by the infinite chain \cite{NT, GGY}
\be
\tr \left( A_{1} B_{1} A_{1} B_{1} \cdots \right) \,.
\ee
The vacuum  preserves an $SU(2|2)$ subgroup of $OSp(2,2|6)$. There are two types
of elementary excitations: ``$A$-particles'', which correspond to
replacing $A_{1}$ by $A_{2}, B_{2}^{\dagger},
(\psi^{\dagger}_{B_{2}})_{\alpha}$; and ``$B$-particles'', which correspond to
replacing $B_{1}$ by $A_{2}^{\dagger}, B_{2},
(\psi^{\dagger}_{A_{2}})_{\alpha}$. We therefore identify the
$B$-particles as charge conjugates of the $A$-particles.

It is convenient to represent
these $A$-particles and $B$-particles by Zamolodchikov-Faddeev
operators $A_{i}^{\dagger}(p)$ and $B_{i}^{\dagger}(p)$ ($i = 1, \ldots, 4$),
respectively. When acting on the vacuum state $|0\rangle$, these
operators create corresponding asymptotic particle states
of momentum $p$ and energy $E$ given by \cite{NT, GGY, GHO}
\be
E = \sqrt{\frac{1}{4}+4 g^{2} \sin^{2}\frac{p}{2}} \,,
\ee
where $g$ is a function of the 't Hooft coupling
\be
g=h(\lambda) \,,
\label{gvalue}
\ee
with $h(\lambda) \sim \lambda$ for small $\lambda$, and
$h(\lambda) \sim \sqrt{\lambda/2}$ for large $\lambda$.

We define the $A$-$A$ $S$-matrix by
\be
A_{i}^{\dagger}(p_{1})\, A_{j}^{\dagger}(p_{2}) =
S_{\ \ \ \ i\, j}^{AA\ i' j'}(p_{1}, p_{2})\,
A_{j'}^{\dagger}(p_{2})\, A_{i'}^{\dagger}(p_{1}) \,.
\ee
The $SU(2|2)$ symmetry implies that, up to a scalar factor, this
$S$-matrix is the same as the one for ${\cal N}=4$ YM theory. Hence,
\be
S^{AA}(p_{1}, p_{2}) = S_{0}(p_{1}, p_{2})\, \widehat S(p_{1}, p_{2}) \,,
\label{SmatrixAA}
\ee
where $\widehat S(p_{1}, p_{2})$ is the $SU(2|2)$ $S$-matrix
\cite{AFZ} with $g$ given by (\ref{gvalue}).  It satisfies the
Yang-Baxter equation, as well as unitarity
\be
\widehat S_{12}(p_{1}, p_{2})\, \widehat S_{21}(p_{2}, p_{1}) = \id
\label{hatSunitarity}
\ee
and the crossing relation \cite{Ja}
\be
\widehat S_{12}^{t_{2}}(p_{1},p_{2})\, C_{2}\, \widehat S_{12}(p_{1}, \bar p_{2})\,
C_{2}^{-1}  =
\widehat S_{12}^{t_{1}}(p_{1},p_{2})\, C_{1}\, \widehat S_{12}(\bar p_{1}, p_{2})\,
C_{1}^{-1} = f(p_{1}, p_{2})\, \id \,,
\label{hatScrossing}
\ee
where $C$ is the charge conjugation matrix, and
\be
f(p_{1},p_{2}) = \frac{\left(\frac{1}{x^{+}_{1}} -
x^{-}_{2}\right)(x^{+}_{1} - x^{+}_{2})}
{\left(\frac{1}{x^{-}_{1}} -
x^{-}_{2}\right)(x^{-}_{1} - x^{+}_{2})} \,.
\label{ffunc}
\ee
As usual,
\be
x^{+}+\frac{1}{x^{+}}-x^{-}-\frac{1}{x^{-}} = \frac{i}{g}\,, \qquad
\frac{x^{+}}{x^{-}} = e^{i p}\,, \quad
\ee
and $x^{\pm}(\bar p) = 1/x^{\pm}(p)$.
Moreover, $S_{0}(p_{1}, p_{2})$ in (\ref{SmatrixAA}) is a scalar factor which is yet to be
determined.

Similarly, we define the $B$-$B$ and $A$-$B$ $S$-matrices by
\be
B_{i}^{\dagger}(p_{1})\, B_{j}^{\dagger}(p_{2}) =
S_{\ \ \ \ i\, j}^{BB\ i' j'}(p_{1}, p_{2})\,
B_{j'}^{\dagger}(p_{2})\, B_{i'}^{\dagger}(p_{1})
\ee
and
\be
A_{i}^{\dagger}(p_{1})\, B_{j}^{\dagger}(p_{2}) =
S_{\ \ \ \ i\, j}^{AB\ i' j'}(p_{1}, p_{2})\,
B_{j'}^{\dagger}(p_{2})\, A_{i'}^{\dagger}(p_{1}) \,,
\ee
respectively. Symmetry considerations suggest that
\be
S^{BB}(p_{1}, p_{2}) &=& S^{AA}(p_{1}, p_{2}) = S_{0}(p_{1}, p_{2})\, \widehat S(p_{1},
p_{2}) \,, \non \\
S^{AB}(p_{1}, p_{2}) &=& S^{BA}(p_{1}, p_{2}) =\tilde S_{0}(p_{1}, p_{2})\, \widehat S(p_{1},
p_{2}) \,,
\label{SmatricesBBAB}
\ee
where $\widehat S(p_{1}, p_{2})$ is the same $SU(2|2)$ matrix as in
(\ref{SmatrixAA}).

We assume that the $S$-matrices satisfy the usual unitarity relation,
\be
S_{12}^{AA}(p_{1}, p_{2})\,  S_{21}^{AA}(p_{2}, p_{1}) =
S_{12}^{AB}(p_{1}, p_{2})\,  S_{21}^{AB}(p_{2}, p_{1}) =\id \,,
\ee
which implies that the scalar factors should obey
\be
S_{0}(p_{1}, p_{2})\, S_{0}(p_{2}, p_{1}) = 1\,, \qquad
\tilde S_{0}(p_{1}, p_{2})\, \tilde S_{0}(p_{2}, p_{1}) = 1 \,.
\label{S0unitarity}
\ee

The identification of the $B$-particles as charge conjugates of the
$A$-particles suggests the crossing relations
\be
S_{12}^{AA\ t_{2}}(p_{1},p_{2})\, C_{2}\, S_{12}^{AB}(p_{1}, \bar p_{2})\,
C_{2}^{-1}  =
S_{12}^{AA\ t_{1}}(p_{1},p_{2})\, C_{1}\, S_{12}^{AB}(\bar p_{1}, p_{2})\,
C_{1}^{-1} = \id \,.
\label{crossing}
\ee
It follows that the scalar factors should satisfy
\be
S_{0}(p_{1},p_{2})\, \tilde S_{0}(p_{1}, \bar p_{2}) =
S_{0}(p_{1},p_{2})\, \tilde S_{0}(\bar p_{1},  p_{2}) =
\frac{1}{f(p_{1},p_{2})} \,.
\label{S0crossing}
\ee
Note that the crossing relations (\ref{crossing}) do {\it not}
relate $S^{AA}$ to itself as in the ${\cal N}=4$ YM theory.

We find that values of the scalar factors which are consistent with the
unitarity and crossing constraints (\ref{S0unitarity}) and
(\ref{S0crossing}) are
\be
S_{0}(p_{1}\,, p_{2}) &=&
\frac{1-\frac{1}{x^{+}_{1}x^{-}_{2}}}{1-\frac{1}{x^{-}_{1}x^{+}_{2}}}
\sigma(p_{1}\,, p_{2}) \,, \non \\
\tilde S_{0}(p_{1}\,, p_{2}) &=&
\frac{x^{-}_{1}-x^{+}_{2}}{x^{+}_{1}-x^{-}_{2}}
\sigma(p_{1}\,, p_{2}) \,,
\label{S0values}
\ee
where $\sigma(p_{1}\,, p_{2})$ is the BES dressing factor \cite{BES},
which has the properties
\be
\sigma(p_{1}, p_{2})\, \sigma(p_{2}, p_{1}) &=& 1 \,, \non \\
\sigma(\bar p_{1}, p_{2})\, \sigma( p_{1},p_{2}) &=&
\frac{x^{-}_{2}}{x^{+}_{2}}\frac{1}{f(p_{1},p_{2})} \,, \qquad
\sigma(p_{1}, \bar p_{2})\, \sigma(p_{1}, p_{2}) =
\frac{x^{+}_{1}}{x^{-}_{1}}\frac{1}{f(p_{1},p_{2})} \,.
\label{sigma}
\ee
As we shall see in the following section, the expressions
(\ref{S0values}) for the scalar factors are crucial for obtaining the
correct BAEs.

\section{Asymptotic Bethe ansatz}\label{sec:ABA}

We now proceed to formulate the asymptotic Bethe ansatz for the ${\cal
N}=6$ CS theory.  The analysis is similar to the one for the ${\cal
N}=4$ YM theory \cite{Be, MM}, and here we follow closely the latter
reference.  We consider a set of $N_{A}$ $A$-particles with momenta
$p^{A}_{i}$ ($i= 1, \ldots, N_{A}$) and $N_{B}$ $B$-particles with
momenta $p^{B}_{i}$ ($i= 1, \ldots, N_{B}$) which are widely
separated on a ring of length $L'$. Quantization conditions for these
momenta follow from imposing periodic boundary conditions on the
wavefunction. Taking a particle with momentum $p^{A}_{k}$ around the
ring leads to the Bethe-Yang equations
\be
e^{-i p^{A}_{k} L'} = \Lambda( \lambda=p^{A}_{k}, \{ p^{A}_{i},
p^{B}_{i} \})\,, \quad k = 1, \ldots, N_{A} \,,
\label{ABetheYang}
\ee
where $\Lambda( \lambda, \{ p^{A}_{i}, p^{B}_{i} \})$ are the
eigenvalues of the transfer matrix \footnote{As emphasized by Martins
and Melo \cite{MM}, in order to properly implement periodic boundary
conditions, it is necessary to use the graded $S$-matrix (which is related to
the non-graded $S$-matrix by the factor ${\cal P}^{(g)}\, {\cal P}$,
where ${\cal P}^{(g)}$ and ${\cal P}$ are the graded and non-graded
permutation matrices, respectively), and take the supertrace (instead
of the ordinary trace) of the monodromy matrix.}${}^{,}$\footnote{Following
\cite{MM}, we denote the spectral parameter of the transfer matrix by $\lambda$, which
should not be confused with the `t Hooft coupling!}
\be
t(\lambda, \{ p^{A}_{i}, p^{B}_{i} \}) = \str_{a}
S^{AA}_{a\, 1}(\lambda, p^{A}_{1}) \cdots
S^{AA}_{a\, N_{A}}(\lambda, p^{A}_{N_{A}})\,
S^{AB}_{a\, N_{A}+1}(\lambda, p^{B}_{1}) \cdots
S^{AB}_{a\, N_{A}+N_{B}}(\lambda, p^{B}_{N_{B}}) \,.
\label{transfer1}
\ee
The order of the particles on the ring is irrelevant.  Indeed,
changing the order of the particles changes the transfer matrix by a
unitary transformation, since the $S$-matrices satisfy the Yang-Baxter
equation.

Using (\ref{SmatrixAA}), (\ref{SmatricesBBAB}), we see that the
transfer matrix can be reexpressed as
\be
t(\lambda, \{ p^{A}_{i}, p^{B}_{i} \}) =
\prod_{i=1}^{N_{A}}S_{0}(\lambda, p^{A}_{i})
\prod_{i=1}^{N_{B}}\tilde S_{0}(\lambda, p^{B}_{i})\,
\widehat t(\lambda, \{ p^{A}_{i}, p^{B}_{i} \}) \,,
\label{transfer2}
\ee
where
\be
\widehat t(\lambda, \{ p^{A}_{i}, p^{B}_{i} \}) = \str_{a}
\widehat S_{a\, 1}(\lambda, p^{A}_{1}) \cdots
\widehat S_{a\, N_{A}}(\lambda, p^{A}_{N_{A}})\,
\widehat S_{a\, N_{A}+1}(\lambda, p^{B}_{1}) \cdots
\widehat S_{a\, N_{A}+N_{B}}(\lambda, p^{B}_{N_{B}}) \,.
\label{hattransfer}\
\ee
It follows that
\be
\Lambda( \lambda, \{ p^{A}_{i}, p^{B}_{i} \}) =
\prod_{i=1}^{N_{A}}S_{0}(\lambda, p^{A}_{i})
\prod_{i=1}^{N_{B}}\tilde S_{0}(\lambda, p^{B}_{i})\,
\widehat \Lambda(\lambda, \{ p^{A}_{i}, p^{B}_{i} \};
\{\lambda_{j}, \mu_{j} \}) \,,
\label{transfereigenvalue}
\ee
where $\widehat \Lambda(\lambda, \{ p^{A}_{i}, p^{B}_{i} \};
\{\lambda_{j}, \mu_{j} \})$ are the eigenvalues
of $\widehat t(\lambda, \{ p^{A}_{i}, p^{B}_{i} \})$, which are
given by \cite{MM}
\be
&& \widehat \Lambda(\lambda, \{ p^{A}_{i}, p^{B}_{i} \}; \{\lambda_{j}, \mu_{j}
\}) =
\prod_{i=1}^{N_{A}}\left[ \frac{x^{+}(\lambda) - x^{-}(p^{A}_{i})}
{x^{-}(\lambda) - x^{+}(p^{A}_{i})}
\frac{\eta(p^{A}_{i})}{\eta(\lambda)} \right]
\prod_{i=1}^{N_{B}}\left[ \frac{x^{+}(\lambda) - x^{-}(p^{B}_{i})}
{x^{-}(\lambda) - x^{+}(p^{B}_{i})}
\frac{\eta(p^{B}_{i})}{\eta(\lambda)} \right] \non \\
&& \quad \times \prod_{j=1}^{m_{1}}\left[\eta(\lambda)
\frac{x^{-}(\lambda)-x^{+}(\lambda_{j})}
{x^{+}(\lambda)-x^{+}(\lambda_{j})}\right]\non \\
&&-
\prod_{i=1}^{N_{A}}\left[ \frac{x^{+}(\lambda) - x^{+}(p^{A}_{i})}
{x^{-}(\lambda) - x^{+}(p^{A}_{i})}
\frac{1}{\eta(\lambda)} \right]
\prod_{i=1}^{N_{B}}\left[ \frac{x^{+}(\lambda) - x^{+}(p^{B}_{i})}
{x^{-}(\lambda) - x^{+}(p^{B}_{i})}
\frac{1}{\eta(\lambda)} \right]\Bigg\{
\prod_{j=1}^{m_{1}}\left[ \eta(\lambda)
\frac{x^{-}(\lambda)-x^{+}(\lambda_{j})}
{x^{+}(\lambda)-x^{+}(\lambda_{j})}\right] \non\\
&& \quad \times
\prod_{l=1}^{m_{2}}
\frac{x^{+}(\lambda) + \frac{1}{x^{+}(\lambda)} - \tilde \mu_{l} +
\frac{i}{2g}}
{x^{+}(\lambda) + \frac{1}{x^{+}(\lambda)} - \tilde \mu_{l} - \frac{i}{2g}}
+\prod_{j=1}^{m_{1}}\left[ \eta(\lambda)
\frac{x^{+}(\lambda_{j})-\frac{1}{x^{+}(\lambda)}}
{x^{+}(\lambda_{j})-\frac{1}{x^{-}(\lambda)}}\right]
\prod_{l=1}^{m_{2}}
\frac{x^{-}(\lambda) + \frac{1}{x^{-}(\lambda)} - \tilde \mu_{l} -
\frac{i}{2g}}
{x^{-}(\lambda) + \frac{1}{x^{-}(\lambda)} - \tilde \mu_{l} + \frac{i}{2g}} \Bigg\} \non \\
&&+
\prod_{i=1}^{N_{A}}\left[ \frac{x^{+}(\lambda) - x^{+}(p^{A}_{i})}
{x^{-}(\lambda) - x^{+}(p^{A}_{i})}
\frac{1-\frac{1}{x^{-}(\lambda) x^{+}(p^{A}_{i})}}
{1-\frac{1}{x^{-}(\lambda) x^{-}(p^{A}_{i})}}
\frac{\eta(p^{A}_{i})}{\eta(\lambda)} \right]
\prod_{i=1}^{N_{B}}\left[ \frac{x^{+}(\lambda) - x^{+}(p^{B}_{i})}
{x^{-}(\lambda) - x^{+}(p^{B}_{i})}
\frac{1-\frac{1}{x^{-}(\lambda) x^{+}(p^{B}_{i})}}
{1-\frac{1}{x^{-}(\lambda) x^{-}(p^{B}_{i})}}
\frac{\eta(p^{B}_{i})}{\eta(\lambda)} \right] \non\\
&& \quad \times
\prod_{j=1}^{m_{1}}\left[ \eta(\lambda)
\frac{x^{+}(\lambda_{j})-\frac{1}{x^{+}(\lambda)}}
{x^{+}(\lambda_{j})-\frac{1}{x^{-}(\lambda)}}\right] \,,
\label{thateigenvalues}
\ee
where $\eta(\lambda) = e^{i \lambda/2}$.
The corresponding BAEs are given by
\be
&& e^{i(P^{A}+P^{B})/2}
\prod_{i=1}^{N_{A}} \frac{x^{+}(\lambda_{j}) - x^{-}(p^{A}_{i})}
{x^{+}(\lambda_{j}) - x^{+}(p^{A}_{i})}
\prod_{i=1}^{N_{B}} \frac{x^{+}(\lambda_{j}) - x^{-}(p^{B}_{i})}
{x^{+}(\lambda_{j}) - x^{+}(p^{B}_{i})}
= \prod_{l=1}^{m_{2}}
\frac{x^{+}(\lambda_{j}) + \frac{1}{x^{+}(\lambda_{j})}
- \tilde \mu_{l} + \frac{i}{2g}}
{x^{+}(\lambda_{j}) + \frac{1}{x^{+}(\lambda_{j})}
- \tilde \mu_{l} - \frac{i}{2g}} \,, \non\\
&& \qquad \qquad \qquad j = 1, \ldots, m_{1} \,,
\non \\
&& \prod_{j=1}^{m_{1}}\frac{\tilde \mu_{l} -
x^{+}(\lambda_{j})-\frac{1}{x^{+}(\lambda_{j})}+\frac{i}{2g}}
{\tilde \mu_{l} -
x^{+}(\lambda_{j})-\frac{1}{x^{+}(\lambda_{j})}-\frac{i}{2g}}
 =
\prod_{k=1 \atop k\ne l}^{m_{2}}
\frac{\tilde \mu_{l} - \tilde \mu_{k} + \frac{i}{g}}
{\tilde \mu_{l} - \tilde \mu_{k} - \frac{i}{g}} \,, \qquad l = 1, \ldots,
m_{2} \,,
\label{BAE}
\ee
where
\be
P^{A}=\sum_{i=1}^{N_{A}}p^{A}_{i} \,, \qquad
P^{B}=\sum_{i=1}^{N_{B}}p^{B}_{i} \,.
\ee

The Bethe-Yang equations for the $A$-particles (\ref{ABetheYang})
therefore take the form
\be
e^{i p^{A}_{k} \left( -L'+ \frac{N^{A}+N^{B}}{2}-\frac{m_{1}}{2}\right)}
&=& e^{i(P^{A}+P^{B})/2}
\prod_{i=1 \atop i \ne k}^{N_{A}}\left[ \frac{x^{+}(p^{A}_{k}) - x^{-}(p^{A}_{i})}
{x^{-}(p^{A}_{k}) - x^{+}(p^{A}_{i})}
 \right]
 \left[\frac{1-\frac{1}{x^{+}(p^{A}_{k}) x^{-}(p^{A}_{i})}}
 {1-\frac{1}{x^{-}(p^{A}_{k}) x^{+}(p^{A}_{i})}}
\sigma(p^{A}_{k}\,, p^{A}_{i}) \right] \non\\
& & \times
\prod_{i=1}^{N_{B}} \sigma(p^{A}_{k}\,, p^{B}_{i})\,
\prod_{j=1}^{m_{1}}\left[
\frac{x^{-}(p^{A}_{k})-x^{+}(\lambda_{j})}
{x^{+}(p^{A}_{k})-x^{+}(\lambda_{j})}\right] \,,
\quad k = 1, \ldots, N_{A} \,.
\label{ABetheYang2}
\ee
In obtaining the result (\ref{ABetheYang2}), we have used the
expressions (\ref{transfereigenvalue}), (\ref{thateigenvalues}) for
the eigenvalues of the transfer matrix (note that only the first term
in (\ref{thateigenvalues}) survives after setting $\lambda =
p^{A}_{k}$), as well as our expressions (\ref{S0values}) for the
scalar factors.
Note that the scalar factor ${\tilde S}_0$ produces a cancelation, so that $A-B$ scattering
in (\ref{ABetheYang2}) is given only by the dressing factor;
and the scalar factor $S_0$ produces a contribution which allows the $A-A$
scattering to be expressed in terms of $u_{4,j}$ (see (\ref{ufourj}) below).

Similarly, the Bethe-Yang equations for the
$B$-particles become
\be
e^{i p^{B}_{k} \left( -L'+ \frac{N^{A}+N^{B}}{2}-\frac{m_{1}}{2}\right)}
&=& e^{i(P^{A}+P^{B})/2}
\prod_{i=1 \atop i \ne k}^{N_{B}}\left[ \frac{x^{+}(p^{B}_{k}) -
x^{-}(p^{B}_{i})}{x^{-}(p^{B}_{k}) - x^{+}(p^{B}_{i})}
 \right]
 \left[\frac{1-\frac{1}{x^{+}(p^{B}_{k}) x^{-}(p^{B}_{i})}}
 {1-\frac{1}{x^{-}(p^{B}_{k}) x^{+}(p^{B}_{i})}}
\sigma(p^{B}_{k}\,, p^{B}_{i}) \right] \non\\
& & \times
\prod_{i=1}^{N_{A}} \sigma(p^{B}_{k}\,, p^{A}_{i})\,
\prod_{j=1}^{m_{1}}\left[
\frac{x^{-}(p^{B}_{k})-x^{+}(\lambda_{j})}
{x^{+}(p^{B}_{k})-x^{+}(\lambda_{j})}\right] \,,
\quad k = 1, \ldots, N_{B} \,,
\label{BBetheYang2}
\ee
since they are formulated in terms of the same transfer matrix
$\widehat t(\lambda, \{ p^{A}_{i}, p^{B}_{i} \})$ (\ref{hattransfer}).

The BAEs (\ref{BAE}) and Bethe-Yang equations
(\ref{ABetheYang2}), (\ref{BBetheYang2}) can be mapped to the all-loop
BAEs of Gromov and Vieira \cite{GV}.  To this end,
we make the following identifications: \footnote{In the ${\cal N}=4$
YM case, a similar set of identifications has been made \cite{MM} in
order to map the $SU(2|2)^{2}$ asymptotic BAEs to
the all-loop BAEs of Beisert and Staudacher
\cite{BS}.}
\be
x^{\pm}(p^{A}_{k}) &=&x^{\pm}_{4, k}\,, \quad k = 1, \ldots,
K_{4} \equiv N_{A} \,, \non \\
x^{\pm}(p^{B}_{k}) &=&x^{\pm}_{{\bar 4}, k}\,, \quad k = 1, \ldots,
K_{\bar 4} \equiv N_{B}\,, \non \\
x^{+}(\lambda_{j}) &=& \frac{1}{x_{1, j}}\,, \quad j = 1, \ldots, K_{1}\,,
\non \\
x^{+}(\lambda_{K_{1}+j}) &=& x_{3, j}\,, \quad j = 1, \ldots,
K_{3}\,, \quad K_{1} + K_{3} \equiv m_{1} \,, \non \\
\tilde \mu_{j} &=& \frac{u_{2, j}}{g}\,, \quad j = 1, \ldots, K_{2}
\equiv m_{2} \,,
\ee
and also define
\be
u_{4, j} = x^{+}_{4, j} + \frac{1}{x^{+}_{4, j}} - \frac{i}{2} =
x^{-}_{4, j} + \frac{1}{x^{-}_{4, j}} + \frac{i}{2}
\label{ufourj}
\ee
(similarly for $u_{{\bar 4}, j}$), and
$u_{i, j} = \frac{1}{g}\left(x_{i, j} + \frac{1}{x_{i, j}}\right)$ for $i=1, 3$.
Indeed, imposing the zero-momentum condition \cite{MZ, GV}
\be
P^{A}+P^{B} = \sum_{j=1}^{K_{4}} p_{4, j} + \sum_{j=1}^{K_{\bar 4}}
p_{{\bar 4}, j} =0 \,,
\ee
the Bethe-Yang equations (\ref{ABetheYang2}) become
\be
e^{i p_{4, k} \left(-L' + \frac{K_{4}+K_{\bar
4}+K_{1}-K_{3}}{2}\right)} &=&
\prod_{j=1 \atop j \ne k}^{K_{4}}
\frac{u_{4, k} - u_{4, j} + i}{u_{4, k} - u_{4, j} - i}\,
\sigma(u_{4, k}, u_{4, j})
\prod_{j=1}^{K_{\bar 4}} \sigma(u_{4, k}, u_{{\bar 4}, j}) \non \\
&& \times
\prod_{j=1}^{K_{1}} \frac{1-\frac{1}{x^{-}_{4, k} x_{1, j}}}
{1-\frac{1}{x^{+}_{4, k} x_{1, j}}}
\prod_{j=1}^{K_{3}} \frac{x^{-}_{4, k} - x_{3, j}}
{x^{+}_{4, k} - x_{3, j}} \,, \quad k = 1, \ldots, K_{4} \,.
\label{ABetheYang3}
\ee
This agrees with the all-loop BAEs for $x^{\pm}_{4, k}$ in
\cite{GV}, provided we identify
\be
L = -L' + \frac{K_{4}+K_{\bar 4}+K_{1}-K_{3}}{2} \,.
\label{LpL}
\ee
If we assume that $L'=-J$ (as in \cite{MM}), then (\ref{LpL}) implies
a relation among $L$, $J$ and the number of various Bethe roots.
Similarly, the Bethe-Yang equations (\ref{BBetheYang2}) give the the
all-loop BAEs for $x^{\pm}_{{\bar 4}, k}$ in
\cite{GV}.

The first set of BAEs in (\ref{BAE}) imply the
following two sets of equations
\be
\prod_{i=1}^{K_{4}}\frac{1-\frac{1}{x_{1, j} x^{-}_{4, i}}}
{1-\frac{1}{x_{1, j} x^{+}_{4, i}}}
\prod_{i=1}^{K_{\bar 4}}\frac{1-\frac{1}{x_{1, j} x^{-}_{{\bar 4}, i}}}
{1-\frac{1}{x_{1, j} x^{+}_{{\bar 4}, i}}} &=&
\prod_{l=1}^{K_{2}}\frac{u_{1, j} - u_{2, l} + \frac{i}{2}}
{u_{1, j} - u_{2, l} - \frac{i}{2}} \,, \quad j = 1, \ldots, K_{1}\,,
\non \\
\prod_{i=1}^{K_{4}}\frac{x_{3, j} -x^{-}_{4, i}}
{x_{3, j} -x^{+}_{4, i}}
\prod_{i=1}^{K_{\bar 4}}\frac{1-\frac{1}{x_{3, j} x^{-}_{{\bar 4}, i}}}
{1-\frac{1}{x_{3, j} x^{+}_{{\bar 4}, i}}} &=&
\prod_{l=1}^{K_{2}}\frac{u_{3, j} - u_{2, l} + \frac{i}{2}}
{u_{3, j} - u_{2, l} - \frac{i}{2}} \,, \quad j = 1, \ldots, K_{3}\,,
\ee
which agree with the first and third BAEs in
\cite{GV}. Finally, the second set of BAEs in (\ref{BAE}) imply
\be
\prod_{j=1 \atop j\ne l}^{K_{2}}
\frac{u_{2, l} - u_{2, j} + i}{u_{2, l} - u_{2, j} - i}
= \prod_{j=1}^{K_{1}}
\frac{u_{2, l} - u_{1, j} + \frac{i}{2}}
{u_{2, l} - u_{1, j} - \frac{i}{2}}
\prod_{j=1}^{K_{3}}
\frac{u_{2, l} - u_{3, j} + \frac{i}{2}}
{u_{2, l} - u_{3, j} - \frac{i}{2}} \,, \quad l = 1, \ldots, K_{2}\,,
\ee
which agrees with the second set of BAEs in
\cite{GV}. In short, the asymptotic BAEs which follow from the
proposed $S$-matrix give the full set of all-loop BAEs in \cite{GV}.

\section{Discussion}\label{sec:discussion}

We have seen that, as in ${\cal N}=4$ YM, $SU(2|2)$ symmetry again
plays an important role in determining the factorizable $S$-matrix of
${\cal N}=6$ CS. The scattering matrices for the two types of
particles are the same, up to the dressing phases which are related by
a new crossing relation.  This relation seems to be weaker than the
one in the $AdS_5/CFT_4$ case, which completely determines the phase
up to the usual CDD ambiguity.  Hence, more analysis is necessary to fix the
dressing phases uniquely.  The solution proposed here is supported by
the all-loop BAEs conjectured in \cite{GV}.

It should be possible to make further checks of our proposal.
Perturbative computations at both small and large values of `t Hooft
coupling could be done.  The classical $S$-matrix can be computed, as
was done for the giant magnon for $AdS_5/CFT_4$ \cite{HM}.  These
results may give some insights into the relation between the parameter
$g$ in the $S$-matrix and the `t Hooft coupling $\lambda$ defined in
(\ref{gvalue}).

One interesting application of the $S$-matrix is to compute
finite-size effects, or the L\"uscher correction.  The result can be
compared with various semi-classical string calculations
\cite{fineff} as has been done for the $AdS_5/CFT_4$
in \cite{AFZi} and with perturbative CS computations along the line of \cite{BajJan}.
The $S$-matrix can also be used to determine the bound
state spectrum, which should be related to the multiplet states
studied in \cite{GGY}.

Our result shows that the $SU(2|2)$-invariant $S$-matrix appears in
both $AdS_5/CFT_4$ and $AdS_4/CFT_3$ dualities.  It would be
interesting to understand this common integrability property at a more
fundamental level.

\section*{Acknowledgments}
This work was performed at the 2008 APCTP Focus Program ``Finite-size
technology in low-dimensional quantum systems (4)''.
We thank the participants, particularly P. Dorey, for discussions.
We also thank M. Martins for helpful correspondence.
This work was supported in part by KRF-2007-313-C00150 (CA) and by the
National Science Foundation under Grants PHY-0244261 and PHY-0554821
(RN).

\end{document}